\documentclass[12pt]{article}

\usepackage[paperheight=11in,paperwidth=21.cm,margin=0.8in,heightrounded]{geometry}

\usepackage{ulem} 

\usepackage{hyperref} 
\hypersetup{colorlinks=true, linkcolor=blue, filecolor=blue, urlcolor=mygreen, citecolor=red}
\usepackage{breakurl} 
\usepackage{times} 

\usepackage{cite}
\usepackage{epsfig}
\usepackage{amsmath}
\usepackage{array}
\usepackage{cancel}
\usepackage{color} 
\definecolor{red}{rgb}{0.9,0,0} 
\definecolor{myblue}{rgb}{0,0.2,0.4} 
\definecolor{mygreen}{rgb}{0.0,0.75,0.0} 

\def\mathswitch#1{\relax\ifmmode#1\else$#1$\fi}
\def\mathswitchr#1{\relax\ifmmode{\mathrm{#1}}\else$\mathrm{#1}$\fi}
\newcommand{\PW}{\mathswitchr W}
\newcommand{\PZ}{\mathswitchr Z}
\newcommand{\PH}{\mathswitchr H}
\newcommand{\Pe}{\mathswitchr e}
\newcommand{\Pb}{\mathswitchr b}
\newcommand{\Pt}{\mathswitchr t}

\newcommand{\Pq}{\mathswitchr q}
\newcommand{\MW}{\mathswitch {M_\PW}}
\newcommand{\MZ}{\mathswitch {M_\PZ}}
\newcommand{\GZ}{\mathswitch {\Gamma_\PZ}}
\newcommand{\MH}{\mathswitch {M_\PH}}

\newcommand{\mt}{\mathswitch {m_\Pt}}

\newcommand{\mz}{\mathswitch {\overline{M}_\PZ}}

\newcommand{\gz}{\mathswitch {\overline{\Gamma}_\PZ}}
\newcommand{\as}{\alpha_{\mathrm s}}
\newcommand{\at}{\alpha_\Pt}
\newcommand{\seff}[1]{\sin^2\theta_{\rm eff}^{\rm #1}}

\newcommand{\gev}{\,\, \mathrm{GeV}}

\newcommand{\re}{\Re e \,}

\newcommand{\OO}{{\mathcal O}}

\newcommand{\mycaption}[1]{\caption{\sl #1}}

\begin{document}
\thispagestyle{empty}
\allowdisplaybreaks

\def\thefootnote{\fnsymbol{footnote}}

\begin{flushright}
DESY 16-119
\\
KW 16-002 
\\
July 2016
\end{flushright}

\vspace{1cm}

\begin{center}

{\Large {\bf The two-loop electroweak bosonic corrections 
to $\seff{b}$}}
\\[3.5em]
{\large
Ievgen Dubovyk$^{1}$, 
Ayres~Freitas$^2$, 
Janusz Gluza$^3$, 
\\[2mm]
Tord Riemann$^{3,4}$, 
Johann Usovitsch$^{5}$
}
\end{center}

\vspace*{1cm}

{\sl \small
\noindent
$^{1}$ 
II. Institut f{\"u}r Theoretische Physik, Universit{\"a}t Hamburg,
22761 Hamburg,  Germany 
\\[1ex]
$^2$ Pittsburgh Particle physics, Astrophysics \& Cosmology Center
(PITT PACC),\\ Department of Physics \& Astronomy, University of Pittsburgh,
Pittsburgh, PA 15260, USA 
\\[1ex]
$^3$ Institute of Physics, University of 
Silesia, 
40007 Katowice, Poland
\\[1ex]
$^{4}$ 
15711 K{\"o}nigs Wusterhausen, Germany
\\[1ex]
$^{5}$ 
Institut f{\"u}r Physik, Humboldt-Universit{\"a}t zu Berlin, 12489 Berlin, Germany}


\vspace*{2.5cm}

\begin{abstract}
The prediction of the effective electroweak mixing angle $\sin^2\theta_{\rm eff}^{\rm b}$ in the Standard Model at two-loop 
accuracy has now been completed by
the first calculation of the bosonic two-loop corrections to the $Z{\bar b}b$ vertex.
Numerical predictions are presented in the form of a fitting formula as function of 
$M_Z, M_W, M_H, m_t$ and $\Delta{\alpha}$, ${\alpha_{\rm s}}$.
For central input values, we obtain a relative correction of
$\Delta\kappa_{\rm b}^{(\alpha^2,\rm bos)} = -0.9855 \times 10^{-4}$, 
amounting to about a quarter of the fermionic corrections, and 
corresponding to $\sin^2\theta_{\rm eff}^{\rm b} = 0.232704$.
The integration of the corresponding two-loop vertex Feynman integrals with up to three dimensionless 
parameters in Minkowskian kinematics has been performed with two approaches: (i)
Sector decomposition, implemented in the  packages {\tt FIESTA\,3} and {\tt
SecDec\,3}, and 
(ii)
Mellin-Barnes representations, implemented in
 {\tt AMBRE\,3/MB} and the new package 
{\tt MBnumerics}.
\end{abstract}

\setcounter{page}{1} 
\setcounter{footnote}{0}

\newpage


\section{Introduction}

This paper reports on the calculation of the
{\it bosonic} $\OO(\alpha^2)$ corrections to $\seff{b}$ and to the $Z\to b\bar{b}$ decay asymmetry parameter $A_\Pb$.
{Here {\it bosonic} refers to corrections from diagrams without
closed fermion loops.}
This completes the calculation of their $\OO(\alpha^2)$
electroweak corrections: The {\it fermionic} two-loop corrections have been given in Ref.~\cite{Awramik:2008gi}.

For the {\it leptonic} $Z$ decay asymmetry parameter,
the calculation of the complete electroweak two-loop corrections was presented in 
\cite{Awramik:2006ar,Awramik:2006uz}.
For the other $Z$-boson parameters 
-- 
$\GZ^{\ell}$, $\GZ^{\nu}$, $\GZ^\Pq$,  $\GZ^\Pb$ --,
and for $A_\Pb$,
the {\it fermionic}  electroweak two-loop corrections have been determined 
\cite{Awramik:2008gi,Freitas:2013dpa,Freitas:2014hra}, but
the {\it bosonic} electroweak two-loop corrections were yet unknown. 

We would like to remind the reader that $e^+e^-$-annihilation into fermion pairs is described by a gauge invariant, unitary and analytic 
scattering amplitude \cite{Stuart:1991xk}:
\begin{align}\label{eq-smatrix}
 {\cal \overline{M}}^0 &\sim \frac{R}{s-{\bar s}_0} +S + (s-{\bar s}_0)~S' + \cdots,
\qquad
{\bar s}_0 =  \mz^2- i \mz \gz .
\end{align}
The proper formalism for its perturbative calculation has been derived in 
\cite{Stuart:1991cc,Veltman:1992tm} and its application at two-loop accuracy is described
in Ref.~\cite{Awramik:2006uz} and 
references therein.
The amplitude \eqref{eq-smatrix} has a Breit-Wigner resonance form
with  fixed mass $\mz$ and width {$\gz$}.
A Breit-Wigner ansatz
with {an} energy-dependent width $\Gamma_\PZ$ as it is used in most experimental analyses leads to a 
numerically different mass $\MZ$, and the two mass definitions can be translated by 
\cite{Bardin:1988xt} 
\begin{equation}
\textstyle
\mz = \MZ\big/\sqrt{1+\Gamma_\PZ^2/\MZ^2}\,, \qquad
\gz = \Gamma_\PZ\big/\sqrt{1+\Gamma_\PZ^2/\MZ^2}\, \label{massrel}
.
\end{equation}
The arguments apply to $\MW$ as well.
While we have used the on-shell
masses $\overline{M}$ in our calculations, the numerical results in
section~\ref{sec:res} are reported in terms of the commonly used masses $M$.

The residue $R$ in \eqref{eq-smatrix} factorizes, in an excellent approximation, into initial- and final state vertex 
form factors, $V^{Ze^+e^-}_\mu$ and 
$V^{Zb\bar{b}}_\nu$, and $Z$-propagator corrections, $R_\PZ^{\mu\nu}$. For this reason, the unfolded $Z$-peak forward-backward asymmetry 
$A^{b\bar{b},0}_{\rm FB}$ and forward-backward left-right asymmetry $A^{b\bar{b},0}_{\rm FB,LR}$ 
can be written, also in an excellent approximation, as
\begin{align}
\label{eq-pseudo-defs}
A^{b\bar{b},0}_{\rm FB} &= \tfrac{3}{4} A_\Pe A_\Pb, &
A^{b\bar{b},0}_{\rm FB,LR} &= \tfrac{3}{4} P_\Pe A_\Pb, &
\end{align}
where $P_\Pe$ is the electron polarization and
\begin{align}
\label{eq-def-ab0}
A_\Pb &= 
\frac{2 ~ {\re} \frac{g_{\rm V}^b}{g_{\rm A}^b}}
{ 1+ \left( {\re} \frac{g_{\rm V}^b}{g_{\rm A}^b} \right)^2}
~~=~~
 \frac{1-4|Q_b|\seff{b}}{1-4|Q_b|\seff{b}+8Q_b^2
 \bigl ( \seff{b} \bigr )^2}
.
\end{align}
The right part of \eqref{eq-def-ab0} follows from the definition
\begin{align}
\seff{b} &= \frac{1}{4|Q_b|}\biggl (1-\re \frac{g_{\rm V}^b}{g_{\rm A}^b} \biggr), 
\label{eq:seff} 
\end{align}
where $Q_b=-1/3$.
Technically, the calculation of $A_\Pb$ rests on the calculation of the vertex form factor
$V^{Zb\bar{b}}_\mu$, whose vector and axial-vector components can be obtained using the projection operations
\begin{align}
g_{\rm V}^b(k^2) &= \frac{1}{2(2-D)k^2} \, {\rm Tr}[\gamma^\mu \, \cancel{p}_1 \,
V^{Zb\bar{b}}_\mu \, \cancel{p}_2], 
\\
g_{\rm A}^b(k^2) &= \frac{1}{2(2-D)k^2} \, {\rm Tr}[\gamma_5 \,
\gamma^\mu \, \cancel{p}_1 \,
V^{Zb\bar{b}}_\mu \, \cancel{p}_2],
\end{align}
where $D=4-2\epsilon$ is the space-time dimension and $p_{1,2}$ are the 
momenta of the
external $b$-quarks, and $k=p_1+p_2$. 
As a result, only scalar integrals remain after projection,
but they may contain non-trivial combinations of scalar products in the numerator.
More specifically, we here calculate the bosonic two-loop contribution to the (complex) ratio
${g_{\rm V}^b(\MZ^2)}/{g_{\rm A}^b(\MZ^2)}$.

The determination of the pseudo-observables in Eq.~\eqref{eq-pseudo-defs} from true observables requires carefully 
written interfaces for the unfolding and subtraction of QED, QCD and box contributions, 
and other contributions not contained in the pseudo-observables;
see Refs.~\cite{Bardin:1999yd,Bardin:1999gt}.
In fact, the interfaces implemented in {\tt ZFITTER} 
\cite{Bardin:1999yd,Arbuzov:2005ma,web-sanc.zfitter:2016} 
have proven to be adequate for an analysis of $Z$-pole pseudo-observables at 
the
$\OO(\alpha^2)$ level \cite{Awramik:2006uz}.\footnote{%
In fitting programs like {\tt Gfitter} 
\cite{Gfitter:20160716} it is assumed that the validity of \eqref{eq-pseudo-defs}
was established by data preparation.
In this respect we would like to mention that a natural language for the unfolding  of measured cross-sections 
into pseudo-observables (not discussed here) and the relation of pseudo-observables to theory predictions has been worked out in the 
S-matrix approach 
\cite{Leike:1991pq,Riemann:1992gv,Riemann:2015wpn}. We will discuss this topic elsewhere.}
The experimental values for $A_\Pb$ and $\seff{b}$ from a global fit to the LEP and SLC data are \cite{ALEPH:2005ablast0,Agashe:2014kda}:
\begin{align}
 A_\Pb &= 0.899 \pm 0.013, & \seff{b} &= 0.281 \pm 0.016
.
\end{align}

\medskip 

A challenge has been the evaluation of two-loop vertex 
integrals in the Minkowskian kinematic region. The vertices 
involve up to three additional mass scales besides $s=\MZ^2$, and many of them also contain ultraviolet
(UV) and infrared (IR) singularities, even though the divergencies cancel in the final result. In general, it is not 
possible to compute all 
integrals analytically with available methods  and tools, but instead one has to resort to numerical
integration strategies. The techniques used in this work are discussed in
section~\ref{sec:2l}. Results for the numerical impact of the new corrections 
are presented in section~\ref{sec:res}, before the summary in
section~\ref{sec:sum}.



\section{Strictly numerical two-loop integration techniques}
\label{sec:2l}
The complete set of two-loop integrals required for this calculation can be divided into several categories. For the 
renormalization counterterms one needs two-loop self-energies with Minkowskian external momentum, $p^2 = M_i^2 + i\varepsilon$, $M_i = 
\MW,\, \MZ$. In addition, there are two-loop vertex integrals with
one non-vanishing external momentum squared, $s = \MZ^2 + i\varepsilon$.
Two-loop self-energy 
integrals and vertex integrals with self-energy subloops have been computed using the dispersion relation techniques described in 
Refs.~\cite{Bauberger:1994by,Bauberger:1994hx,Awramik:2006uz}.
The remaining two-loop  vertex integrals with triangle subloops amount to
some 700 integrals, 
with tensor rank $R\leq 3$, Minkowskian external momentum,
and up 
to three dimensionless parameters per integral, from the set 
$M_i^2/\MZ^2$, where $\MH^2, \MW^2, \mt^2$, besides $M_i^2=\MZ^2+i\varepsilon$. 
The aim is an accuracy of eight significant digits, to be obtained with two completely independent calculations. 

A variety of integrals were calculated already for the  leptonic $Z$ boson asymmetry parameter $A_\Pe$. Here, up to two 
 dimensionless parameters had to be treated \cite{Awramik:2006ar,Awramik:2006uz,Czakon:2006pf}. 
In view of the larger number of scales encountered in the $Zb\bar{b}$ vertex, and also aiming at
comparably simple and semi-automatic algorithms with easy 
re-use, a fully numerical strategy was applied here. 

No reduction to a minimal set of master integrals (MIs) was attempted, except
for simple cancellations of numerator and denominator terms. There are several
reasons; none of them is stringent. One might perform a standard reduction to
MIs, which could reduce the number of integrals by about a factor of ten. From
the point of view of performance of the project as a whole, this is no important
gain in efficiency, because the time of calculating the integrals is not a
limiting factor. On the other hand, for cases with many different mass scales,
coefficient terms in integral reductions can become very large, which makes this
approach cumbersome from a technical point of view. Furthermore, using the
program KIRA~\cite{Kira:2016tobesub}, we observed that the numerical treatment
of the coefficient terms becomes difficult for some integrals with propagators
of mass $\MZ$ at the kinematical point $s = \MZ^2 + i\varepsilon$. At the same
time, we know that a number of MIs will remain to be evaluated numerically. The
techniques developed for these can relatively easily be applied to the complete
set of (unreduced) integrals, thus obviat- ing the need for integral reductions.
Finally, our goal was to create a self-contained general-purpose numerical
package, see Ref.~\cite{Dubovyk:2016-tobepublished} for more details. Reductions
and partial analytical solutions are difficult to integrate into this.

As was mentioned above,
individual integrals
will contain both UV and/or soft and collinear divergencies.
{We have employed} two techniques with an automatic control of these divergencies: sector 
decomposition (SD) and 
Mellin Barnes (MB) representations. 

It is essential that the numerical methods work sufficiently stable for Minkowskian kinematics.
{For sector decomposition \cite{Hepp:1966eg,Binoth:2000ps} this 
can be achieved through a complex contour deformation of the Feynman parameter integrals, as implemented in}
the publicly available 
packages {\tt FIESTA 3} \cite{Smirnov:2013eza} since 2013 and {\tt SecDec 3}  
\cite{Borowka:2015mxa} since 2015.
Nevertheless, we observe serious convergence problems for some of our integrals.
As a second, independent method we chose the representation of Feynman integrals by Mellin-Barnes integrals 
\cite{Usyukina:1975yg,Smirnov:1999gc,Tausk:1999vh}.
The MB method has been well developed in recent years and there are useful software packages available at the {\tt MBtools} 
webpage in the {\tt hepforge} archive \cite{mbtools}: 
{\tt MB} \cite{Czakon:2005rk}, 
{\tt MBresolve} \cite{Smirnov:2009up}, 
{\tt AMBRE 1} \cite{Gluza:2007rt} and 
{\tt barnesroutines} (D.~Kosower).
Further, one may use 
{\tt PlanarityTest} \cite{Bielas:2013rja}, {\tt AMBRE 2} \cite{Gluza:2010rn} and {\tt AMBRE 3} \cite{Dubovyk:2015yba}, as well as {\tt 
MBsums} 
\cite{Ochman:2015fho}, which are available from the {\tt AMBRE} webpage \cite{Katowice-CAS:2007}.
{For our purposes, we have derived MB representations with {\tt AMBRE}
and used the package} {\tt MB}, aided by {\tt MBresolve} 
and 
{\tt barnesroutines}, for a derivation of an expansion in terms of $\epsilon=(4-d)/2$.
In particular, {\tt AMBRE~2} has been employed for {\it planar} and {\tt AMBRE 3} for {\it non-planar} topologies, using {\tt 
PlanarityTest} for the automatic identification of the planarity status.
{For the numerical treatment of massive MB integrals with Minkowskian kinematics,}
the package {\tt MBnumerics} is being developed since 2015 \cite{Usovitsch:201606}.
For the final numerical integration, it calls the {\tt CUHRE} routine of the {\tt CUBA} library 
\cite{Hahn:2004fe,Hahn:2014fua}.
Some general features of both {\tt AMBRE 3} and {\tt MBnumerics/MB} have been described recently 
\cite{Gluza:ll2016,Riemann:ll2016}.
{For cross-checks,
a variety of integrals was also calculated with {\tt NICODEMOS} 
\cite{Freitas:2012iu} and dispersion relation techniques \cite{Awramik:2006uz}.}
For one-scale integrals, we could make several comparisons with existing analytical and semi-analytical results 
\cite{Fleischer:1998nb,Aglietti:2003yc,Aglietti:2004tq,Aglietti:2007as}.
A comprehensive description of our numerical package \cite{Usovitsch:201606}, including a discussion of the 
numerical derivation of the $Z\to b {\bar b}$ integrals  
will be given elsewhere \cite{Dubovyk:2016-tobepublished}.

\subsection{Using sector decomposition}
\begin{figure}[t!]
\begin{center}
\includegraphics[width=0.24\linewidth]{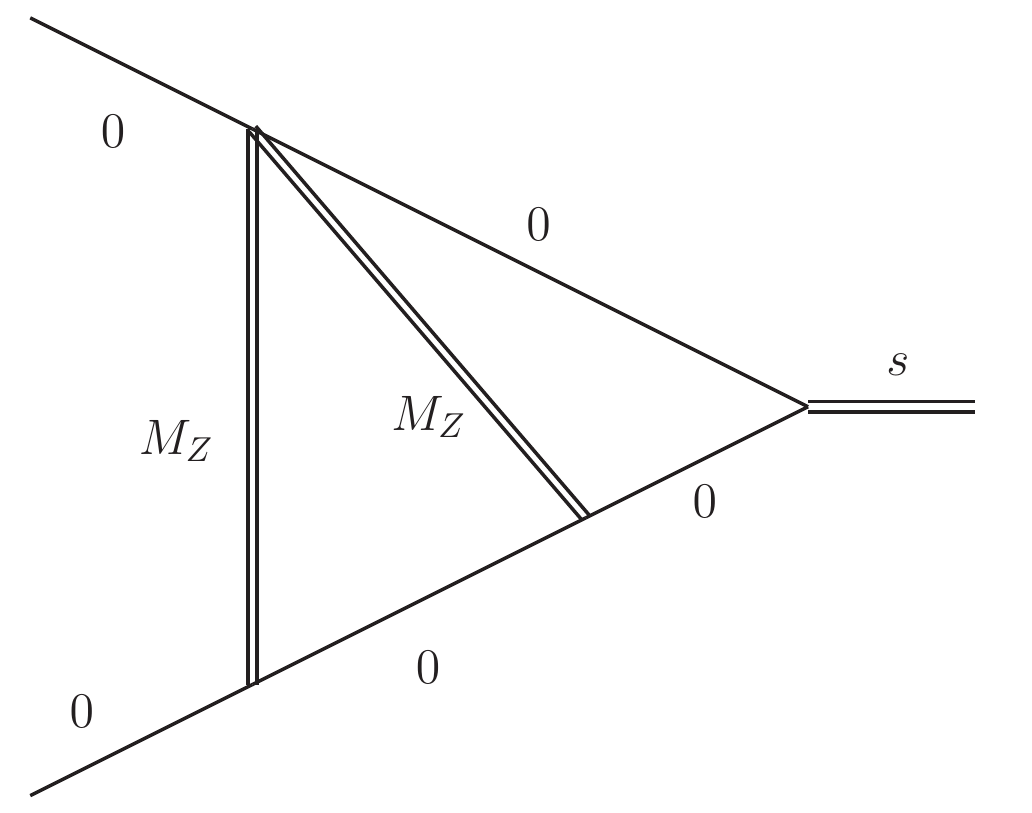}
~
\includegraphics[width=0.22\linewidth]{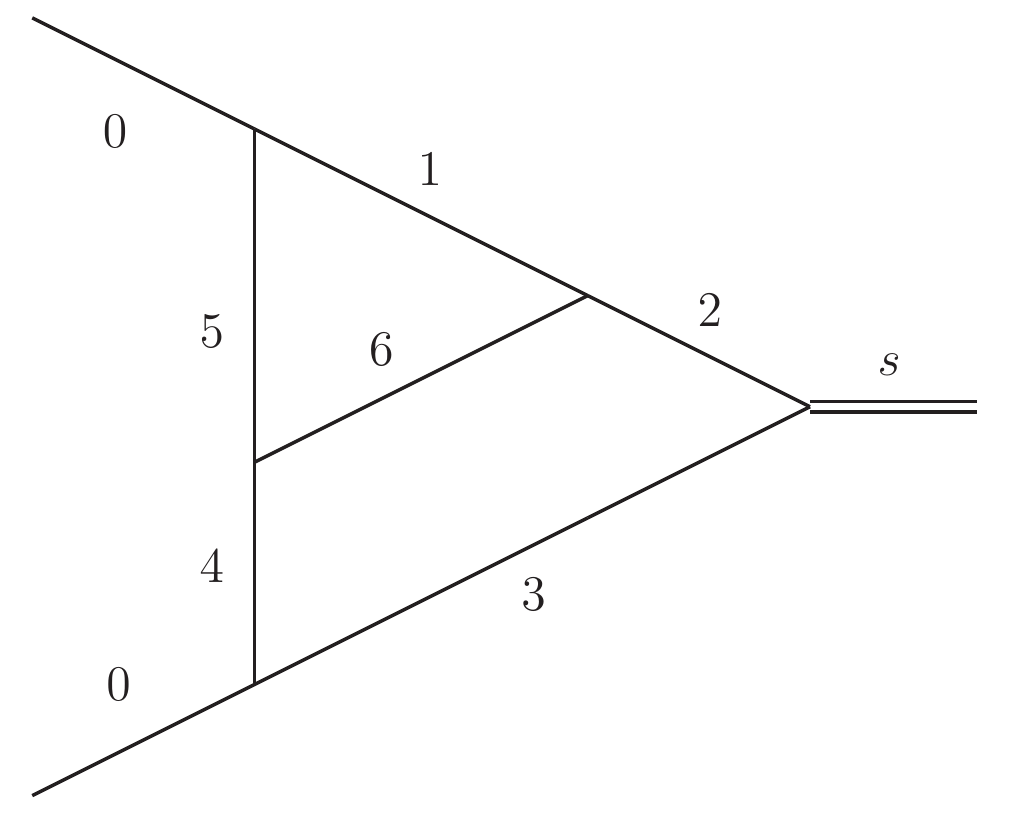}
~
\includegraphics[width=0.22\linewidth]{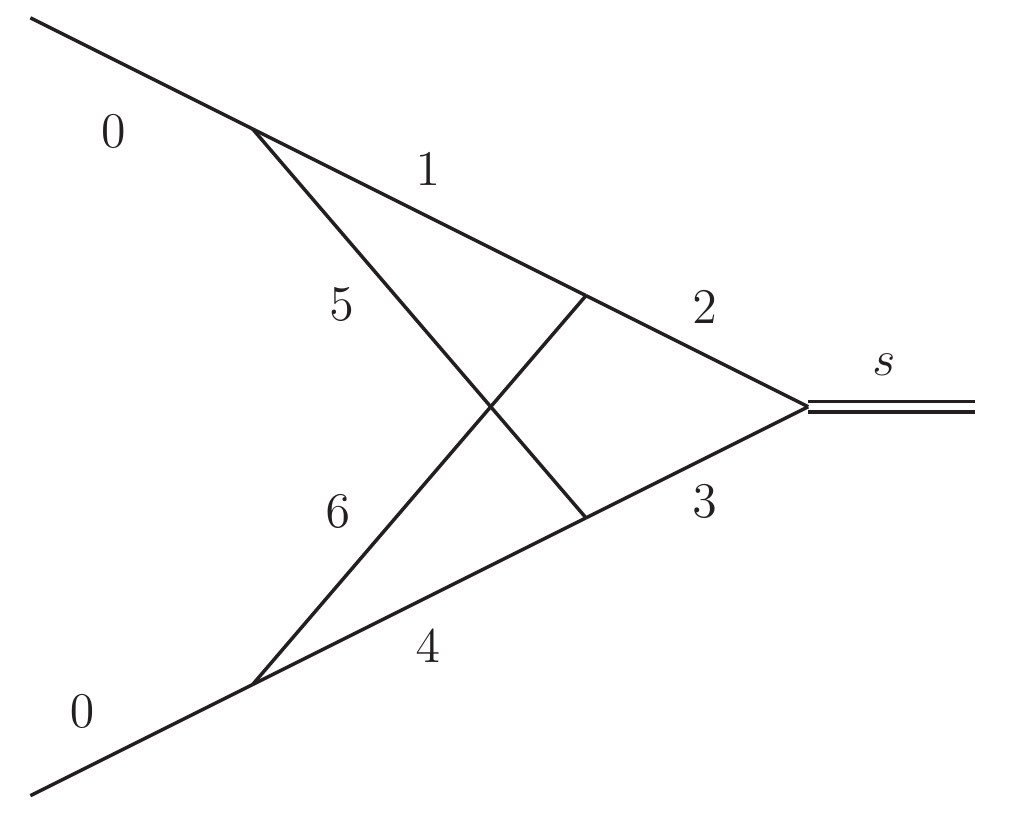}
~
\includegraphics[width=0.22\linewidth]{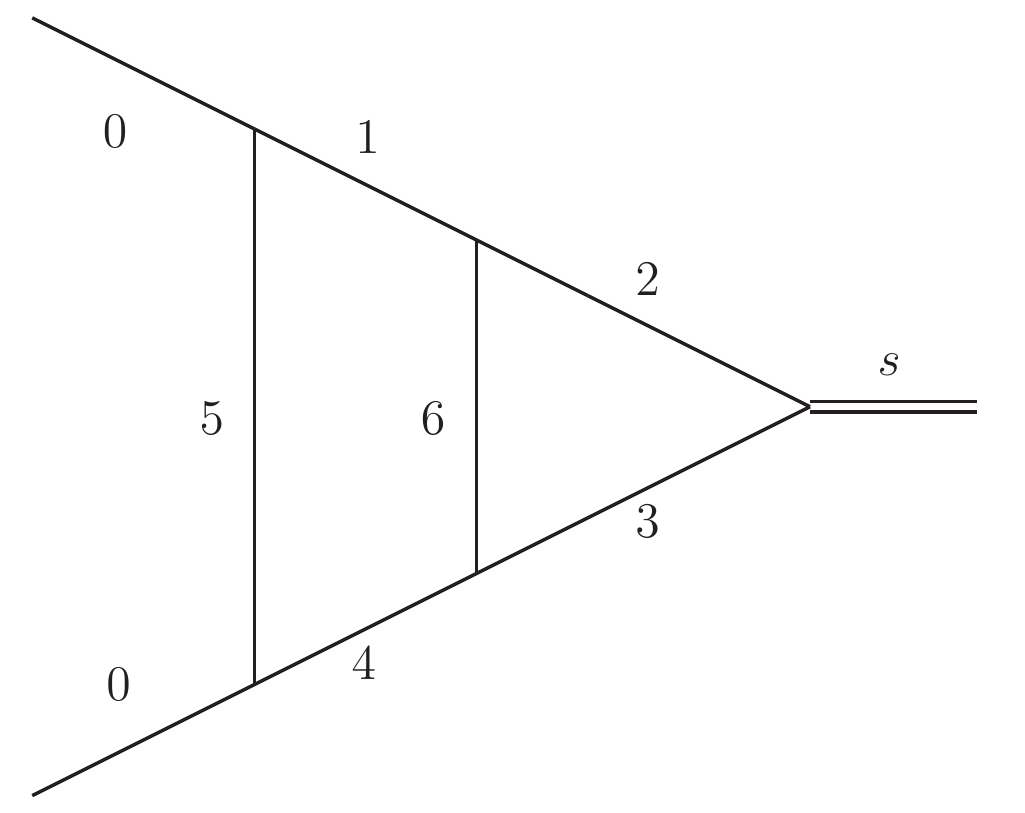}\\
\makebox[0.24\linewidth]{(a)}
~
\makebox[0.22\linewidth]{(b)}
~
\makebox[0.22\linewidth]{(c)}
~
\makebox[0.22\linewidth]{(d)}
\end{center}
\vspace{-2ex}
\mycaption{\label{fig-samples}
Samples of Feynman integral topologies for the $Z{\bar b}b$ vertex.}
\end{figure}

For \textit{Euclidean kinematics} 
all the needed
integrals can be evaluated straightforwardly with sector decomposition, using the packages {\tt FIESTA} and {\tt SecDec}. 
One obtains Feynman parameter integrals with 4 or 5 dimensions.
For \textit{Minkowskian kinematics} the numerical SD method still works well for most of the integrals,
although there were some problematic cases:
\begin{itemize}
\item 
For 16 single-scale six-propagator integrals with one massive line and $s=\MZ^2$, no result at all was obtained with sector decomposition: see 
Fig.~\ref{fig-samples} (b) with $m_4=\MZ$, (c) with $m_1=\MZ$, (d)  with $m_5=\MZ$.
The corresponding MB-representations are at most 3-dimensional.
\item 
For 12 single-scale six-propagator integrals with two massive lines and $s=\MZ^2$,
results with only few significant digits were achieved with sector decomposition: see Fig.~\ref{fig-samples} (b) with $m_1=m_4=\MZ$, (c) 
with $m_1=m_4=\MZ$, (d)  with $m_5=m_6=\MZ$.
The corresponding MB-representations are at most 4-dimensional.
\item 
For 26 
 planar integrals with zero threshold and $s=\MZ^2$, the number of 
integration points had to be increased up to several millions to reach a numerical accuracy of few digits with sector decomposition: see 
Fig.~\ref{fig-samples} (b) with $m_4=\MZ$ or $m_4=0$ and $m_1=\MW,\mt$ and $m_5=m_6=\mt,\MW$, (d) with 
$m_1=\MZ$, where $m_2=\MW,\mt$ and $m_3=m_6=\mt,\MW$. 
The corresponding MB-representations are at most 4-dimensional.
\item
 For 8 
 planar integrals with zero threshold and $s=\MZ^2$, the number of 
integration points had to be increased to about 80 millions in order to determine 
six significant digits with sector decomposition: 
see Fig.~\ref{fig-samples} (d) with $m_5=m_6=\MW,\mt$ and $m_1=m_2=\mt,\MW$, and also with 
$m_5=\MZ$ and $m_6=\MW,\mt$ and $m_2=m_3=\mt,\MW$.
The corresponding MB-representations are at most 5-dimensional.
\end{itemize}
With our implementation of the alternative Mellin-Barnes method,
 at least 8 significant digits were achieved for all 
integrals in this list,
with exclusion of the last item  where we obtain an accuracy of 6 digits.

\subsection{Using the MBtools suite}
The number of dimensions of the Mellin-Barnes integrals increases with the number of mass scales and the complexity of the integral 
topology.
 {\tt AMBRE 2.1} and {\tt AMBRE 3} find the lowest dimensionality of the MB integrals to be solved
\cite{Dubovyk:2015yba,Gluza:ll2016}.\footnote{In some cases, lower dimensionality may be obtained when the 
integrands are allowed to contain 
hypergeometric functions in addition to $\Gamma$-functions and their derivatives; see Eq.~(20) of \cite{Freitas:2010nx}.}
The largest number of MB dimensions encountered here is eight: 
for the constant terms of the non-planar integrals shown in Fig.~\ref{fig-samples} (c) with $m_2=\MZ, m_3=\MH$
and $m_1=m_6=\mt, \MW, m_4=m_5=\MW,\mt$.
For \textit{Euclidean kinematics} we could confirm, sometimes with a lower accuracy, that 
all the MB representations are correct.

{Now let us turn to the treatment of \textit{Minkowskian kinematics} with the MB method.}
From a technical point of view, one has to integrate over products and ratios of
$\Gamma$-functions and their derivatives, multiplied by products of  
terms like 
$[-(s+i\varepsilon)/M^2]^{f(z_i)}$. {Here $f(z_i)$ are linear functions of the} MB integration variables $z_i$, 
{which are parameterized as}
$z_i=x_i+i t_i$, where the $x_i$ are fixed and $t_i \in 
(-\infty,+\infty)$.
{The integrands are rapidly varying and, for Minkowskian kinematics, }
may be highly oscillating and  slowly vanishing at infinity.
There is a 
variety of methods to improve the convergence of Minkowskian MB integrals.
We mention here those which {proved} to be most efficient, but refer for details to the literature 
\cite{Gluza:ll2016,Dubovyk:2016-tobepublished}:
\begin{itemize}
 \item 
{\it Integrand mappings.} 
Before applying a standard integration routine like {\tt CUHRE}, we found a tangent mapping
to be efficient, $t_i \to  1/\tan(-\pi t_i)$,
combined with calculating $\exp[\sum_i\ln(\Gamma_i)]$ rather than  
the product $\Pi_i\,\Gamma_i$.
\item
{\it Contour rotations.} {The transformation} $z_i=x_i+it_i \to {\bar z}_i = x_i+(\theta_i+i)t_i$ may improve the damping 
{of oscillatory terms} 
like $[-(s+i\varepsilon)/M^2]^{f(z_i)}$
at infinity. 
For multi-dimensional MB integrals, 
one may try to perform ``synchronized'' rotations using a universal parameter $\theta_i \equiv \theta$, in order to 
avoid
crossing of poles by the contour 
change \cite{Freitas:2010nx}.
However, for 
single-scale integrals, which depend only on 
$[-(s+i\varepsilon)/\MZ^2]^{f(z_i)} = (-1-i\varepsilon)^{f(z_i)}$, the contour rotation will not improve the 
behaviour at infinity.
\item {\it Contour shifts.}
It proved to be extremely efficient to make use of a well-known property of 
$\Gamma$-function:
At the negative axis between the pole positions, its value becomes smaller when
the function is evaluated at an argument further way from the origin. If a pole
gets crossed by an argument shift,
one has to add the corresponding residue which by itself is also an 
integral, but will have a dimension 
less than the original one.
Doing this several times, with several integration variables, 
the original MB integral gets replaced by several lower-dimensional integrals which may be  easier to calculate, plus the original one 
with shifted integration path.
The resulting smaller contribution of the original integral to the net result has the effect that its poor knowledge 
gets numerically less important.
In effect, the procedure consists of a summing over a finite number of residues with a controlled 
remainder. Shifts of the integration contours were proposed first in Ref.~\cite{Anastasiou:2005cb}.
\end{itemize}
The package {\tt MBnumerics} is yet under development. Currently, it can treat Minkowskian MB integrals with up to 
four dimensions and Euclidean ones with up to five dimensions with a good precision, and with more dimensions at reduced precision.
The package is currently limited to integrals with few different mass scales.

There is the opportunity of internal cross-checks
by integral reductions with the package {\tt KIRA} \cite{Kira:2016tobesub}, followed by a numerical evaluation with MBnumerics.m.
This procedure was used for few integrals where a second calculation was difficult,
and an accuracy of at least 6 digits was reached in these cases for the second calculations.
Further improvement is possible, but the  result is more than sufficient for the purposes of the present calculation.

We will complete this section with a few numerical examples.

\medskip 

The planar IR divergent integral 
\ref{fig-samples} (d) 
depends on 
$s=\MZ^2+i\varepsilon$ and on $\MW$, $\mt$, i.e. on two dimensionless parameters.
The MB representation is three-dimensional, and with {\tt AMBRE/\linebreak[3]MB/\linebreak[3]MBnumerics/CUHRE} we got after 43 minutes 
computer time:\footnote{Here and elsewhere only significant digits are shown.}
\begin{eqnarray}
  I_{1d,\rm MB} &=&  1.541402128186602 + 0.248804198197504 ~i
\\\nonumber 
&&+~ \frac{1}{\epsilon}(0.12361459942846659 - 1.0610332704387688 ~i)
\\\nonumber 
&&+~
  \frac{1}{\epsilon^2}(-0.33773737955057970 + 3.6 \times 10^{-17} ~i)
\end{eqnarray}
Using 24 hours on the same computer, we obtained with {\tt SecDec}:
\begin{eqnarray}
  I_{1d,\rm SD} &=& 1.541 + 0.2487~i + \frac{1}{\epsilon}(0.123615 - 1.06103~i)
\\\nonumber &&
+~
  \frac{1}{\epsilon^2}(-0.3377373796 - 5\times 10^{-10}~i) 
.
\\\nonumber
\end{eqnarray}

With the MB method, we solved all the 100 integrals which depend on only one parameter,
 $\frac{s}{\MZ^2} = 1+i\varepsilon$.
The one-scale integrals  have up to four MB dimensions and were the testing ground during the development of {\tt MBnumerics} 
\cite{Usovitsch:201606}. 
One may calculate all these integrals using the results of 
\cite{Fleischer:1998nb,Aglietti:2003yc,Aglietti:2004tq,Aglietti:2007as}. 
Unfortunately, the authors did not provide a ready-to-use {implementation for numerical evaluation}.
So it was more efficient for us to apply our numerical packages and to perform  additional checks for some selected cases.
We like to  mention here two examples.

\medskip

Integral (a) of Fig.~\ref{fig-samples} is a finite planar five-propagator Feynman integral.
We derived a representation with several  $2$-dimensional MB integrals, requiring about 300 seconds for an 
accuracy 
of 14 digits with 
{\tt AMBRE 2/MB/MBnumerics}, and got
\begin{eqnarray}
\label{eq-f017withmbnumerics}
I_{1a,\rm MB} &=& 
-2.1375883865794   - i~ 3.0210985089304 .
\end{eqnarray}

The integral is analytically known from \cite{Aglietti:2004tq} as a combination of 
generalized harmonic polylogarithms:
$\MZ^2 F_0^{17} = \zeta_2 H(0,1,x)-2H(0,1,0,-1,x)+2H(0,r,r,0,x)$. 
We derived its value at $x=-s/\MZ^2=-1-i\epsilon$: 
\begin{eqnarray}
 \MZ^2 F_0^{17} &=&
\zeta_2 ~ \textrm{Li}_2(-1-i\epsilon)
- ~ \frac{3}{20}\zeta_2^2
+ 2 
\bigl\{
 \pi^4/50 + 2 \ln^4(1/2 + \sqrt{5}/2)
\\ \nonumber
&&
+ ~
 \ln[1/2 + \sqrt{5}/2]^2 \ln[3/2 + \sqrt{5}/2]^2 
-
 1/24 \ln^3(3/2 + \sqrt{5}/2) \ln(2889 + 1292 \sqrt{5})
\\ \nonumber && + ~i~ \bigl[
-(4/3) \pi \ln^3(1/2 + \sqrt{5}/2) + 
 2 \pi \ln^2(1/2 + \sqrt{5}/2) \ln(3/2 + \sqrt{5}/2) 
\\  \nonumber
&&
+ 
 (\pi/6) \ln(7/2 - 3 \sqrt{5}/2) \ln[2/(3 + \sqrt{5})]^2 + 
 (2/5) \pi \zeta_3
\bigr] 
\bigr\}
\\\nonumber
&=& 2.13758838657949792824410730067 + i~ 3.02109850893046314176278063460.
\end{eqnarray}
The difference in sign compared to \eqref{eq-f017withmbnumerics} is due to different metrics. 

\medskip

The non-planar finite integral (c) of Fig.~\ref{fig-samples} with two massive lines 
may be written as a 
four-dimensional MB integral and was solved with a series of contour shifts. 
The needed computer time to determine 11 digits amounts to 
few minutes:
\begin{eqnarray}
I_{1c,\rm MB}  &=& -1.2116223301  + 4.9954503192~i
. 
\end{eqnarray}
The integral is known from \cite{Aglietti:2007as}. It is one of three master integrals, 
calculated by solving a system of differential 
equations (DEQ) numerically. At $s/\MZ^2=1+i\varepsilon$ it is:
\begin{eqnarray}
I_{1c,\rm DEQ}= 16 \times a_0 &=& -1.211622330156316914 + 
4.99545031920035447~i
. 
\end{eqnarray}
To reach an accuracy 
better than the 11 digits shown above with {\tt MBnumerics} would require some effort, but is feasible. 

\medskip

The $1/\epsilon^2$ poles have been verified to cancel analytically and 
numerically for $g_{\rm V}^b(\MZ^2)/g_{\rm A}^b(\MZ^2)$, with more 
than 12 digits precision.
The cancellation of the  $1/\epsilon$ poles has been checked 
numerically with 8 digits precision. 
For the finite part, we obtain a net precision 
of better than 7 digits, which is more than sufficient for practical purposes.


\section{Results}
\label{sec:res}

\begin{table}[tb]
\renewcommand{\arraystretch}{1.2}
\begin{center}
\begin{tabular}{|lll|}
\hline
Parameter & Value & Range \\
\hline \hline
$\MZ$ & 91.1876 GeV & $\pm 0.0042$~GeV \\
$\Gamma_\PZ$ & 2.4952 GeV & \\
$\MW$ & 80.385 GeV & $\pm 0.030$~GeV \\
$\Gamma_\PW$ & 2.085 GeV & \\
$\MH$ & 125.1 GeV & $\pm 5.0$~GeV \\
$\mt$ & 173.2 GeV & $\pm 4.0$~GeV \\
$\as$ & 0.1184 & $\pm 0.0050$ \\
$\Delta\alpha$ & $0.0590$ & $\pm 0.0005$ \\
\hline
\end{tabular}
\end{center}
\vspace{-2ex}
\mycaption{Reference values used in the numerical analysis,
from Ref.~\cite{Agashe:2014kda}.
\label{tab:input}}
\end{table}

The Standard Model prediction for the effective weak mixing angle can be written
as
\begin{equation}
\seff{b} = \Bigl (1-\frac{\MW^2}{\MZ^2}\Bigr )(1+\Delta\kappa_\Pb),
\end{equation}
where $\Delta\kappa_\Pb$ contains the contributions from radiative corrections.
For the numerical analysis, the inputs listed in Tab.~\ref{tab:input} have
been used as default values. With these values, the {\it bosonic} electroweak two-loop
corrections amount to
\begin{equation}
\Delta\kappa_\Pb^{(\alpha^2,\rm bos)} = -0.9855 \times 10^{-4}.
\end{equation}
This result can be compared with the already known corrections:
one-loop contributions \cite{Akhundov:1985fc,Beenakker:1988pv}, {\it fermionic} electroweak
two-loop corrections 
\cite{Awramik:2008gi},
$\OO(\alpha\as)$ QCD corrections 
\cite{Djouadi:1987gn,Djouadi:1987di,Kniehl:1989yc,Kniehl:1991gu,Djouadi:1993ss},
\cite{Fleischer:1992fq,Buchalla:1992zm,Degrassi:1993ij,Chetyrkin:1993jp,Czarnecki:1996ei,Harlander:1997zb},
and partial higher-order corrections of orders
$\OO(\at\as^2)$ 
\cite{Avdeev:1994db,Chetyrkin:1995ix},
$\OO(\at\as^3)$ 
\cite{Schroder:2005db,Chetyrkin:2006bj,Boughezal:2006xk},
$\OO(\alpha^2\at)$ and
$\OO(\at^3)$ 
\cite{vanderBij:2000cg,Faisst:2003px}. The numerical values for the
corresponding contributions are listed in Tab.~\ref{tab:orders}.

As evident from the table, the new bosonic two-loop result is about a factor of
four smaller, but of similar order of magnitude, as the {\it fermionic}
electroweak two-loop corrections \cite{Awramik:2008gi}.%
\footnote{Of course,
this statement is dependent on the fact that we employ the on-shell renormalization
scheme and use $\MW$ as an input parameter at this point.}
\begin{table}[tb]
\renewcommand{\arraystretch}{1.2}
\begin{center}
\begin{tabular}{|lr|}
\hline
Order & Value [$10^{-4}$] \\
\hline \hline
$\alpha$ & 468.945 \\
$\alpha\as$ & $-42.655$ \\
$\at\as^2$ & $-7.074$ \\
$\at\as^3$ & $-1.196$ \\
\hline
\end{tabular}
\begin{tabular}{|lr|}
\hline
Order & Value [$10^{-4}$] \\
\hline \hline
$\at^2\as$ & 1.362 \\
$\at^3$ & 0.123 \\
$\alpha^2_{\rm ferm}$ & 3.866 \\
$\alpha^2_{\rm bos}$ & $-0.986$ \\
\hline
\end{tabular}
\end{center}
\vspace{-2ex}
\mycaption{Comparison of different orders of radiative corrections to
$\Delta\kappa_\Pb$, using the input parameters in Tab.~\ref{tab:input}.
\label{tab:orders}}
\end{table}

For varying input parameters, the new result is best expressed in terms of a
simple fitting formula,
\begin{equation}
\label{eq:formkap}
\Delta\kappa_\Pb^{(\alpha^2,\rm bos)} = k_0 + k_1 c_\PH + k_2  c_\Pt + 
k_3  c_\Pt^2 + k_4  c_\PH c_\Pt + k_5 c_\PW,
\end{equation}
with
\begin{equation}
\begin{aligned}
c_\PH &= \log\left(\frac{\MH}{\MZ} \times \frac{91.1876\gev}{125.1 \gev}\right), &
c_\Pt &= \left(\frac{\mt}{\MZ}\times\frac{91.1876\gev}{173.2 \gev}\right)^2 -1,
\\ &&
c_\PW &= \left(\frac{\MW}{\MZ} \times \frac{91.1876\gev}{80.385 \gev}\right)^2 -1.
\end{aligned}
\end{equation}
Fitting this formula to the full numerical result, the coefficients are obtained
as
\begin{equation}
\begin{aligned}
k_0 &= -0.98605 \times 10^{-4}, & k_1 &= 0.3342 \times 10^{-4}, & k_2 &= 1.3882
\times 10^{-4},  \\
k_3 &= -1.7497 \times 10^{-4}, & k_4 &= -0.4934\times 10^{-4}, & k_5 &= -9.930
\times 10^{-4}.
\end{aligned}
\end{equation}
This parameterization reproduces the full calculation with average and maximal
deviations of $5\times 10^{-8}$ and $1.2\times 10^{-7}$, respectively, for the input parameter
ranges indicated in Tab.~\ref{tab:input}.

Combining this result with the already known corrections (see above)
the currently most precise prediction for $\seff{b}$ is
obtained. Additionally, one free parameter can be eliminated by using the
Standard Model prediction of $\MW$ from the Fermi constant $G_\mu$. The
$W$-boson mass has been calculated previously including
the same perturbative higher order contributions as listed above 
\cite{Awramik:2003rn}. 
To a very good approximation, this result can be written as
\begin{equation}
\label{eq:formsw}
\seff{b} = s_0 + d_1 L_H + d_2  L_H^2 + d_3 \Delta_\alpha +
d_4  \Delta_t + d_5  \Delta_t^2 + d_6 \Delta_tL_H + d_7 \Delta_{\as} + 
d_8 \Delta_t\Delta_{\as} + d_9 \Delta_Z
\end{equation}
with
\begin{equation}
\begin{aligned}
L_H &= \log\left(\frac{\MH}{125.7 \gev}\right), &
\Delta_t &= \left(\frac{\mt}{173.2 \gev}\right)^2 -1, &
\Delta_Z &= \frac{\MZ}{91.1876 \gev} -1, \\
\Delta_\alpha &= \frac{\Delta\alpha}{0.0059} -1, &
\Delta_{\as} &= \frac{\as}{0.1184} -1.
\end{aligned}
\end{equation}
Here $\Delta\alpha$ is the shift of the electromagnetic fine structure constant
due to light fermion loops between the scales $q^2=0$ and $\MZ^2$. The best-fit
numerical values for the coefficients are given by
\begin{equation}
\begin{aligned}
s_0 &= 0.232704, & d_1 &= 4.723\times 10^{-4}, & d_2 &= 1.97\times 10^{-4}, &
d_3 &= 2.07 \times 10^{-2}, \\
&& d_4 &= -9.733\times 10^{-4}, & d_5 &= 3.93\times 10^{-4}, & 
d_6 &= -1.38\times 10^{-4}, \\
&& d_7 &= 2.42\times 10^{-4}, & d_8 &= -8.10\times 10^{-4}, &
d_9 &= -0.664.
\end{aligned}
\end{equation}
With these values, the formula \eqref{eq:formsw} approximates the full result
with average and maximal
deviations of $2\times 10^{-7}$ and $1.3\times 10^{-6}$, respectively, within the ranges in
Tab.~\ref{tab:input}.



\section{Summary}
\label{sec:sum}

The determination of the electroweak two-loop corrections to $A_\Pb$ and $\seff{b}$ has been 
completed.
We have shown by explicit calculation that the numerical result for their bosonic corrections 
 is expectedly small 
compared to the presently available experimental accuracy.
However, the anticipated measurements at a future accelerator of the {ILC/FCC-ee/CEPC generation aim for an accuracy comparable to 
electroweak two-loop effects
\cite{Baer:2013cma,Gomez-Ceballos:2013zzn,CEPC-SPPCStudyGroup:2015csa}.}

Applications related to Drell-Yan processes at the LHC are also of the single-particle resonance type and may be envisaged with 
the technique developed here. 
The numerical packages {\tt FIESTA}, {\tt SecDec} and the {\tt MBtools} suite with the new packages {\tt AMBRE 3} and {\tt MBnumerics} will 
be sufficient, in combination, for calculating the 
whole class of  massive two-loop self-energy and vertex integrals in the Standard Model and beyond. 


\section*{Acknowledgments}

We would like to thank Peter Marquard for discussions and Peter Uwer and his group ``Phenomenology of Elementary 
Particle Physics beyond the Standard Model'' at Humboldt-Universit\"{a}t zu Berlin for providing computer resources.

The work of \textit{I.D.}\ is supported by a research grant of Deutscher Akademischer Austauschdienst (DAAD) and by 
Deutsches 
Elektronensychrotron DESY.
The work of \textit{J.G.}\ is supported by the Polish National Science Centre (NCN) under the Grant Agreement No. DEC-2013/11/B/ST2/04023.
The work of \textit{A.F.}\ is supported in part by the U.S.\ National Science Foundation under grant PHY-1519175.
{The work of \textit{T.R.}\ is supported in part by an
Alexander von Humboldt Polish Honorary Research Fellowship.
The work of \textit{J.U.}\ is supported by Graduiertenkolleg 1504 ``Masse, Spektrum, Symmetrie`` of Deutsche Forschungsgemeinschaft (DFG).}

\textit{A.F.} gratefully acknowledges the hospitality of the Kavli Institute for
Theoretical Physics China during the final stages of this project.


\small

\end{document}